\documentclass[conference]{IEEEtran}
\IEEEoverridecommandlockouts
\usepackage{cite}
\usepackage{amsmath,amssymb,amsfonts}
\usepackage{algorithmic}
\usepackage{graphicx}
\usepackage{textcomp}
\usepackage{subfigure}
\usepackage{caption} 
\usepackage{bm}
\usepackage[ruled,vlined]{algorithm2e}
\usepackage{xcolor}
\def\BibTeX{{\rm B\kern-.05em{\sc i\kern-.025em b}\kern-.08em
    T\kern-.1667em\lower.7ex\hbox{E}\kern-.125emX}}

\def\bb{\mathbf{b}}

\def\bb{\mathbf{b}}

\def\bC{\mathbf{C}}

\def\tt{\textnormal}

\def\tai{DeepIC}

\begin{document}

\title{DeepIC: Coding for Interference Channels via Deep Learning\\
}

\author{\IEEEauthorblockN{Karl Chahine}
\IEEEauthorblockA{\textit{Electrical and Computer Engineering} \\
\textit{University of Texas at Austin}\\
\url{karlchahine@utexas.edu}}
\and
\IEEEauthorblockN{Nanyang Ye}
\IEEEauthorblockA{\textit{Computer Science} \\ 
\textit{Shanghai Jiao Tong University}\\
\url{ynylincoln@sjtu.edu.cn}}
\and
\IEEEauthorblockN{Hyeji Kim}
\IEEEauthorblockA{\textit{Electrical and Computer Engineering} \\
\textit{University of Texas at Austin}\\
\url{hyeji.kim@austin.utexas.edu}}
}

\maketitle

\begin{abstract}
The two-user interference channel is a model for multi one-to-one communications, where two transmitters wish to communicate with their corresponding receivers via a shared wireless medium. Two most common and simple coding schemes are time division (TD) and treating interference as noise (TIN). Interestingly, it is shown that there exists an asymptotic scheme, called Han-Kobayashi scheme, that performs better than TD and TIN. However, Han-Kobayashi scheme has impractically high complexity and is designed for asymptotic settings, which leads to a gap between information theory and practice. 

In this paper, we focus on designing practical codes for interference channels. As it is challenging to analytically design practical codes with feasible complexity, 
we apply deep learning to learn codes for interference channels. We demonstrate that DeepIC, a convolutional neural network-based code with an iterative decoder, outperforms TD and TIN by a significant margin for two-user additive white Gaussian noise channels with moderate amount of interference.
\end{abstract}


\section{Introduction}
\label{sec:Introduction}
\begin{figure}[!ht]
    \centering
    \includegraphics[width=0.48\textwidth]{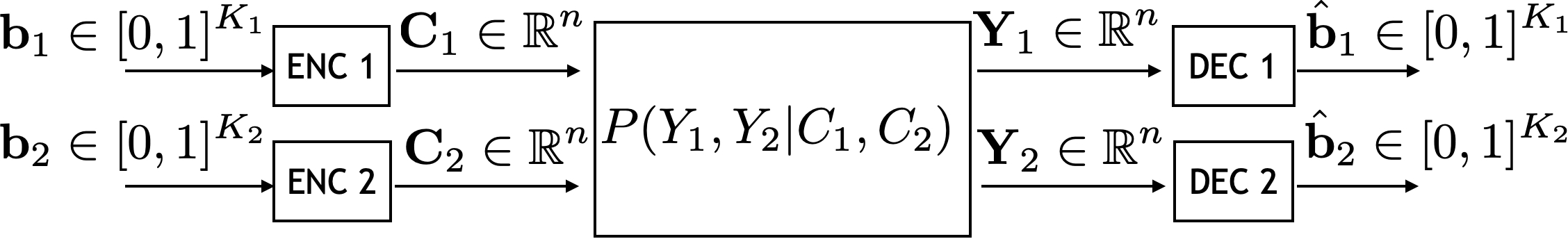}
    \caption{Two-user Interference Channels}
    \label{fig:ic2users}
\end{figure}

The two-user interference channel is a fundamental building block of wireless networks, where two pairs of transmitters and receivers share the communication medium as depicted in Figure~\ref{fig:ic2users}. 
We assume that the $i$-th sender wishes to communicate a bit sequence $\bb_i \in [0,1]^{K_i}$ to the $i$-th receiver, for $i=1,2$. The $i$-th sender maps the message $\bb_i \in [0,1]^{K_i}$ to a length-$n$ sequence $\bC_i \in \mathbb{R}^n$ for $i=1,2$. The capacity region of the two-user interference channel is defined as the set of achievable rate pairs $(R_1,R_2): = (K_1/n, K_2/n)$ such that the probability of error $P((\bb_1, \bb_2) \neq (\hat{\bb}_1, \hat{\bb}_2)) \rightarrow 0$ as $(K_1,K_2,n) \to \infty$ for some encoders and decoders. The capacity region is not known in general and there are inner and outer bounds that coincide for several special cases~\cite{el_gamal_kim_2011,Tse2008,Kobayashi1981}, 
when the interference is very weak or strong. 

Moreover, practical coding schemes for finite $(K_1,K_2,n)$ are vastly missing. Two commonly used schemes are time-division (TD) and treating-interference-as-noise (TIN), both of which are an application of codes designed for point-to-point channels. TD orthogonalizes the transmission of two users and is commonly used in practice. TIN treats the interference as noise and is known to be optimal for weak interference channels. These two are extreme schemes, and a natural question is `can we design a code that is more reliable than TD and TIN for channels with moderate interference?'  
It is challenging to analytically design practical coding schemes with feasible complexity that minimize the probability of errors, however. Hence, in our approach, we apply deep learning to \emph{learn} reliable codes for interference channels. 

Over the past few years, deep learning based approach for designing channel codes and decoders has demonstrated tremendous success for various communication channels~\cite{o2017introduction,nachmani2018deep,Elkelesh2019polar,  
kim2018deepcode,viterbinet}. 
Several work also explored applying deep learning to design modulation for interference channels~\cite{o2017introduction,Wu2020,mishra2021}.~\cite{o2017introduction} demonstrates that feedforward networks can outperform uncoded QAM with time-sharing for very short blocklengths (eg: $4$ bits). ~\cite{Wu2020} proposes an adaptive deep learning algorithm for multi-user symmetric interference channels and show that their algorithm outperforms the conventional system using $PSK$ or $QAM$.~\cite{mishra2021} shows that for interference channels with more than two users, interference alignment schemes can be improved by deep learning.  

Despite the differences, all of the existing work focuses on learning modulation schemes with a very short blocklength. Hence, a very important question remains unanswered. Can we apply deep learning to invent new coding schemes for interference channels that have some memory and work well for longer blocklengths at the order of 10s and 100s? There are several challenges in learning a code with reasonably long blocklengths for interference channels (and any channel in general). First, the number of messages increases exponentially in blocklengths, which implies that we cannot train a code with all possible messages (for blocklength 100, there are $2^{100}$ possible messages). Naturally, strong generalization becomes necessary. Second, learning a code is a non-convex problem, and there tends to be several local optima, requiring a carefully devised optimization strategy. Last, for interference channels, we have to learn a \emph{pair} of codes, which makes the learning problem more challenging. 
%
%

In this paper, we demonstrate DeepIC, a first neural network based code designed for reasonably long blocklengths (eg: $K=100$) that outperforms the state-of-the-art by a large margin. We show that carefully designed neural architectures and training methodology allow us to address the challenges associated with learning codes and achieve a high reliability of DeepIC. 
%
%
Our main contributions are as follows. 

\begin{itemize}
    \item We introduce DeepIC, which consists of two  Convolutional Neural Network (CNN)-based encoders and two CNN-based iterative decoders, designed for two-user interference channels, and the training methodology for DeepIC. 
    \item We show that DeepIC achieves a significant reliability improvement for one hundred coding bits ($K=100$) over the baselines such as TD and TIN, for AWGN channels with moderate interference. DeepIC also outperforms the learning-based methods that directly generalize the state-of-the-art schemes in~\cite{o2017introduction}. 
    \item Through the interpretation analysis, we empirically show that DeepIC has a longer memory (up to around 10) than the direct application of well-known neural architectures. 
\end{itemize}

The rest of the paper is organized as follows. In Section~\ref{sec:setup}, we introduce the problem statement and present a preview of the results. In Section~\ref{sec:Arch}, we provide the details of DeepIC including its neural architecture and training. In Section~\ref{sec:results}, we present experiment results for evaluating the performance of DeepIC as well as for interpreting DeepIC. 

\section{Problem statement}
\label{sec:setup}

\begin{figure}[!htb]
    \centering
    \includegraphics[width=0.49\textwidth]{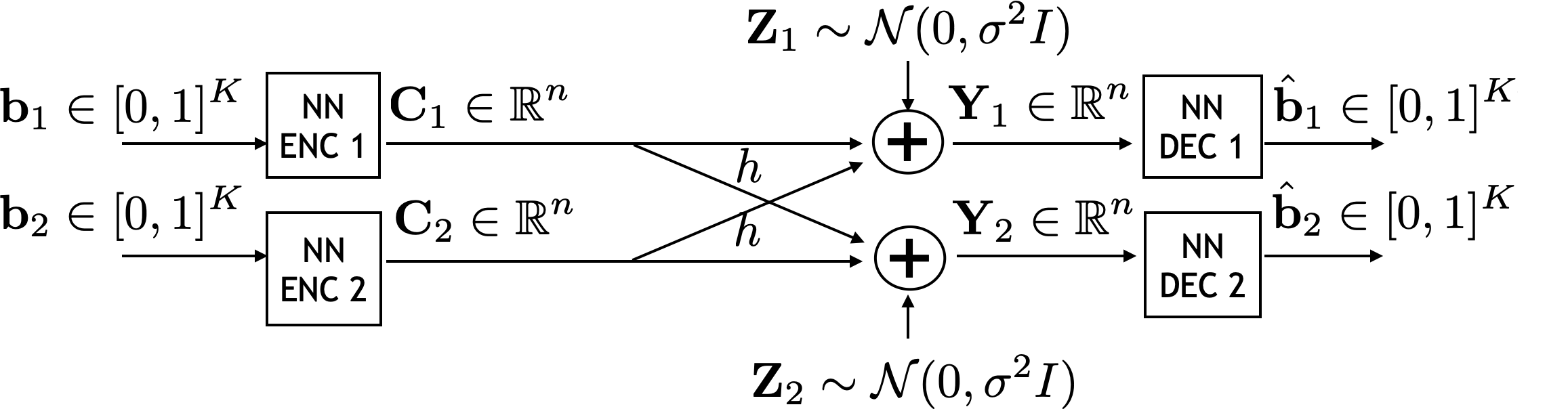}
    \caption{We focus on learning Rate-1/3 reliable codes for 2-user symmetric AWGN interference channels. } 
    \label{fig:Gic}
\end{figure}

We focus on symmetric AWGN interference channels, i.e., $Y_1 = C_1 + h C_2 + Z_1$ and $Y_2 = h C_1 + C_2 + Z_2$, where $Z_1, Z_2 \sim \mathcal{N}(0,\sigma^2)$ as depicted in Figure~\ref{fig:Gic}. 
For concreteness, we focus on the rate-1/3 setting with $K=100$ and $n=300$ for both users. We consider various SNRs, defined as SNR $= -10\log_{10}(\sigma^2)$, and various degree of interference $h=0.3, 0.6, 0.8$. 
Our goal is to design reliable codes that minimize the bit error rate (BER), defined as BER $ = \frac{1}{K}\sum_{i=1}^{K}
P(\hat{b}_{k} \neq b_{k})$. 

%





\subsection{Preview of Results}

\begin{figure}[!htb]
    \centering
    \includegraphics[width=.35\textwidth]{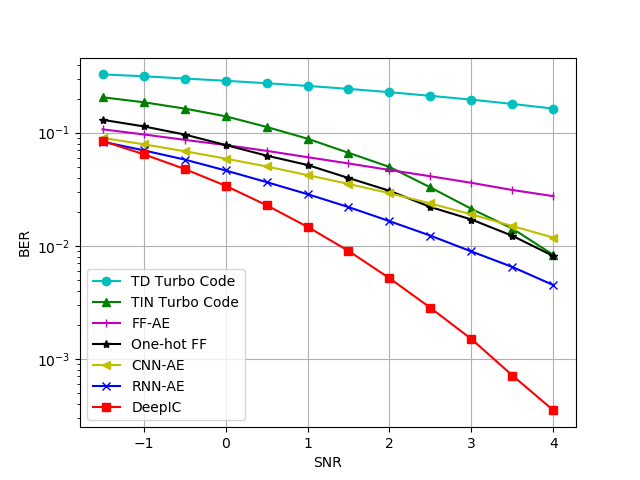}
    \caption{BER (averaged over two users) vs. SNR for h=0.8 and blocklength $K=100$}
    \label{Fig:prev}
\end{figure}

A natural strategy to create a code for interference channels is to utilize the well-known neural networks, such as feedforward networks (FF), CNNs, and Recurrent Neural Networks (RNN). However, in our experiments, we observe that this strategy by itself is not sufficient to achieve a noticeable improvement over the the two baseline schemes, time-division (TD) and treating-interference-as-noise (TIN), as shown in Figure~\ref{Fig:prev}. 
DeepIC, our CNN-based neural codes with an iterative CNN decoding and well-thought-out training, significantly outperforms all the baselines.

\section{Architecture and Training of DeepIC}
\label{sec:Arch}
In this section, we present the \emph{architecture} and the \emph{training methodology} of \tai\ for two-user Gaussian interference channels, which are crucial in achieving the improved reliability. 

\subsection{\textbf {Architecture}}
We construct an autoencoder network for the two-user interference channel by concatenating two encoder-decoder pairs; we replace the encoders and decoders in Figure~\ref{fig:ic2users} by neural networks. 

\begin{figure}[!htb]
    \centering
    \includegraphics[width=.35\textwidth]{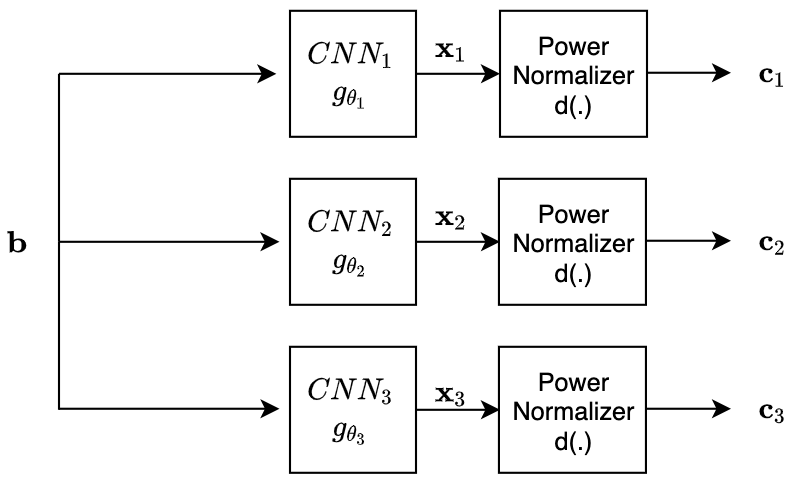}
    \caption{\tai\ Encoder Architecture}
    \label{Fig:enc}
\end{figure}

\noindent \textbf{Encoder architecture:} Each encoder consists of three learnable blocks of 1-D CNNs placed in parallel, followed by a power normalizing layer. As shown in Figure~\ref{Fig:enc}, each message sequence $\textbf{b}$ is encoded into a concatenation of three sequences: $\bm{x_{1}}=g_{\theta_{1}}($\textbf{b}$)$, $\bm{x_{2}}=g_{\theta_{2}}($\textbf{b}$)$, $\bm{x_{3}}=g_{\theta_{3}}(\textbf{b})$, where $g_{\theta_{l}}($\textbf{b}$)$ is the function representing the block CNN$_{l}$ parametrized by the weights $\theta_{l}$, $l \in \{1,2,3\}$. While two encoders share the same encoder architecture, they do not share the weights. 

\begin{figure*}[!ht]
    \centering
    \includegraphics[width=0.7\textwidth]{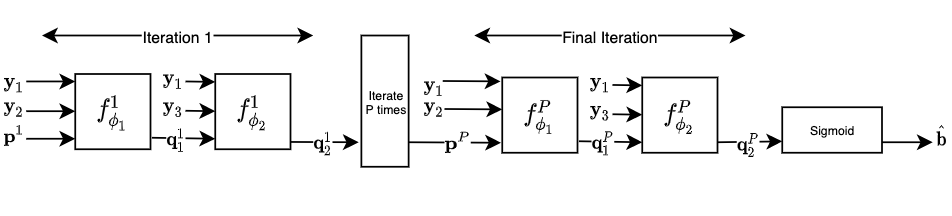}
    \caption{\tai\ Iterative Decoder Architecture}
    \label{fig:dec}
    \vspace{-0.8em}
\end{figure*}

\noindent The power normalizing block $d(.)$ is implemented by estimating the first and second moments of each batch of size $J$. More specifically, for training example $j$, $\bm{b}^{j}$ is encoded into $\bm{x}^{j}$, where $\bm{x}^{j}$ is the concatenation of $\bm{x}_{1}^{j}$, $\bm{x}_{2}^{j}$ and $\bm{x}_{3}^{j}$. Each power normalizer then performs the operation $\bm{c}_{i}^{j}=d(\bm{x}_{i}^{j})$. This is equivalent to calculating $\bm{c}_{i}^{j}=\frac{\bm{x}_{i}^{j}-\bm{\mu_{i}}}{\bm{\sigma_{i}}}$, where $\bm{\mu_{i}} = \frac{1}{J}\sum_{j=1}^{J}\bm{x}_{i}^{j}$ and $\bm{\sigma_{i}}=\sqrt{\frac{1}{J}\sum_{j=1}^{J}(\bm{x}_{i}^{j}-\bm{\mu_{i}})^2}$ are the batch mean and standard deviation, respectively.

\noindent \textbf{Decoder architecture:} Inspired by the dynamic programming decoder, we let the decoder to update the belief in an iterative manner, as depicted in Figure~\ref{fig:dec}. Let $\bm{y}_{1}$, $\bm{y}_{2}$ and $\bm{y}_{3}$ be the noisy versions of 
$\bm{c}_{1}$, $\bm{c}_{2}$ and $\bm{c}_{3}$ respectively. The decoder goes through multiple iterations. Each iteration makes use of two sequential blocks of 1-D CNN layers, $f_{\phi_{1}}$ and $f_{\phi_{2}}$, where $f_{\phi_{l}}$ is the function representing the CNN block parametrized by the weights $\phi_{l}$, $l \in \{1,2\}$. As shown in Figure~\ref{fig:dec}, the first decoder $f_{\phi_{1}}$ takes as input the noisy signals $\bm{y}_{1}$ and $\bm{y}_{2}$, as well as a prior $\bm{p}$ of shape (K,F), where F is the information  feature size (We let $F=5$). The output of the first decoder is the posterior $\bm{q}_{1}$ with same shape (K,F). The second decoder $f_{\phi_{2}}$ takes as input the signals $\bm{y}_{1}$, $\bm{y}_{3}$ and the posterior $\bm{q}_{1}$ to produce the second posterior $\bm{q}_{2}$. The first iteration takes 0 as a prior. At the last iteration, the output posterior's shape is (K,1), and is passed through a sigmoid function to estimate to message bits: $\hat{\bm{b}}=sigmoid(\bm{q}_{2})$.

\subsection{\textbf {Training Methodology}}

The training algorithm used for \tai\ is shown in Algorithm 1. 
The main differences from conventional training are as follows:
\begin{itemize}
    \item At every epoch, the users' encoders and decoders are trained separately. The number of iterations for the encoders and decoders are $T_{enc}$ and $T_{dec}$ respectively. This prevents the training from converging to a local optimum \cite{aoudia2018end}.
    \item Empirically, our results showed that using different training noise levels for the encoders ($\sigma_{enc}$) and decoders ($\sigma_{dec}$) resulted in better performance.
    \item The used batch size was very large (500). This is important to average out the channel noise effects.
    \item Let $L_{1}$ and $L_{2}$ be the binary cross-entropy (BCE) losses of user 1 and user 2 respectively. More specifically, let $\bm{b_{1}}$, $\bm{z_{1}}$ be the message sequence and noise vector of user 1 and $\bm{b_{2}}$, $\bm{z_{2}}$ the message sequence and noise vector of user 2. We compute $L_{1}=BCE(\bm{b_{1}},f_{1}(g_{1}(\bm{b_{1}})+h*g_{2}(\bm{b_{2}})+\bm{z_{1}}))$ and $L_{2}=BCE(\bm{b_{2}},f_{2}(g_{2}(\bm{b_{2}})+h*g_{1}(\bm{b_{1}})+\bm{z_{2}}))$, where $g_{1}$, $f_{1}$ are the encoder and decoder of user 1, and $g_{2}$, $f_{2}$ are the encoder and decoder of user 2.
    In our simulations, we found that evaluating a weighted sum of the losses $L=\alpha*L_{1} + \beta*L{2}$ results in a fair and optimal training for the two users. Note that $\alpha + \beta = 1$, their initial value is 0.5, and they are continuously updated throughout the training. The proposed method is similar as the one described in \cite{o2017introduction}, and is shown in Algorithm 1.
\end{itemize}

\begin{algorithm}
    \label{alg:\tai}
    \SetAlgoLined
    \DontPrintSemicolon
    Inputs: Number of Epochs E, Batch Size J, Encoder Training Steps $T_{enc}$, Decoder Training Steps $T_{dec}$, Encoder Training noise $\sigma_{enc}$, Decoder Training noise $\sigma_{dec}$, $\alpha=\beta=0.5$ (initial values)\\
    \For{$i \leq E$}{
        \For{$k \leq T_{enc}$}{
            Generate examples $\bm{b_{1}}$, noise $\bm{z_{1}}$ $\sim N(0,\sigma_{enc}^2)$\;
            Generate examples $\bm{b_{2}}$, noise $\bm{z_{2}}$ $\sim N(0,\sigma_{enc}^2)$\;
            Compute $L_{1}$, $L_{2}$\ using $\bm{b_{1}}$, $\bm{b_{2}}$, $\bm{z_{1}}$, $\bm{z_{2}}$\;
            Compute $L=\alpha*L_{1}+\beta*L_{2}$\;
            Train the encoders $g_{\theta}$ through back-propagation\;
            Update $\alpha=\frac{L_{1}}{L{1}+L_{2}}$\; 
            Update $\beta=\frac{L_{2}}{L{1}+L_{2}}$ \;
        }

        \For{$k \leq T_{dec}$}{
            Generate examples $\bm{b_{1}}$, noise $\bm{z_{1}}$ $\sim N(0,\sigma_{dec}^2)$\;
            Generate examples $\bm{b_{2}}$, noise $\bm{z_{2}}$ $\sim N(0,\sigma_{dec}^2)$\;
            Compute $L_{1}$, $L_{2}$\ using $\bm{b_{1}}$, $\bm{b_{2}}$, $\bm{z_{1}}$, $\bm{z_{2}}$\;
            Compute $L=\alpha*L_{1}+\beta*L_{2}$\;
            Train the decoders $f_{\phi}$ through back-propagation\;
            Update $\alpha=\frac{L_{1}}{L{1}+L_{2}}$\; 
            Update $\beta=\frac{L_{2}}{L{1}+L_{2}}$ \;
        }
        
    }
    
    \caption{\tai\ pseudo-code}
\end{algorithm}

\noindent Some critical hyperparameters used in our simulations are listed in Table 1.

\begin{table}[h!]
\centering
\begin{tabular}{|c|c|} 
 \hline
 \textbf{Parameter} & \textbf{Assignment} \\ [0.5ex] 
 \hline\hline
    h                       & $\{0.3,0.6,0.8\}$            \\
    Encoders                & 2 layers 1-D CNN          \\
    Decoders                & 5 layers 1-D CNN          \\
    Info Feature Size F     & 5            \\
    Encoder Training Steps $T_{enc}$      & 100            \\
    Decoder Training Steps $T_{dec}$      & 500            \\
    Decoder Iterations      & 6                         \\
    Optimizer               & Adam with initial learning rate 0.0001\\
    Loss                    & Binary Cross-Entropy \\ 
    Number of Epochs E       & 100 \\ [1ex]
 \hline
\end{tabular}
\caption{Hyperparameters of \tai}
\label{Table 1}
\end{table}

\section{Experiment Results}
\label{sec:results}

In this section, we present a set of experiment results and provide intuition on \tai\ 's performance. Throughout this section, BER refers to the average BER of users 1 and 2.

\subsection{\textbf {Performance of \tai\ }}

\begin{figure}[!htb]
    \centering
    \includegraphics[width=.4\textwidth]{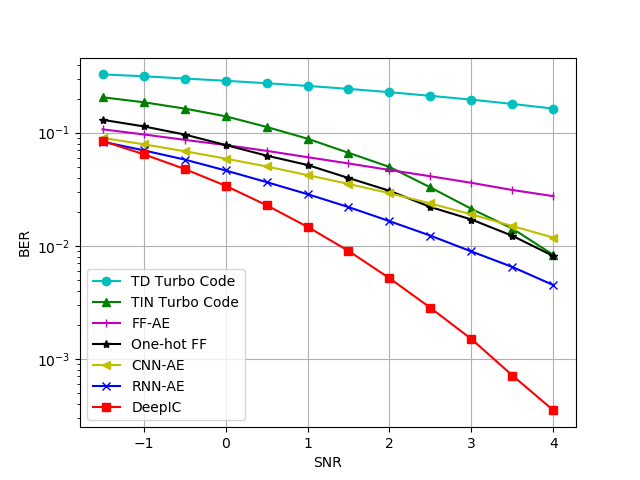}
    \caption{BER performance for h=0.8 and Block Length=100 }
    \label{Fig:fig1}
\end{figure}

To measure performance, we plot the BER of DeepIC as a function of SNR for $h=0.8$ in Figure~\ref{Fig:fig1}. 
As we mentioned in Section~\ref{sec:setup} (A), 
\tai\  shows a clearly better performance when compared to TD and TIN, both of which were implemented using classic Turbo codes. TIN uses a rate-1/3 Turbo code with BPSK, and TD uses a rate-1/3 Turbo code with 4-PAM as in~\cite{o2017introduction},~\cite{Wu2020}.
In addition, we compared \tai\ 's performance with a CNN autoencoder (CNN-AE), a RNN autoencoder (RNN-AE), FeedForward with one-hot encoding (One-hot FF), and a FeedForward autoencoder (FF-AE). \tai\ 's superior performance is coupled to its (a) complex architecture, and (b) personalized training, inducing memory-dependent codewords. This aspect is discussed in the next sub-section.

\subsection{\textbf {Block Length Coding Gain}}

\begin{figure}[!htb]
    \centering
    \includegraphics[width=.4\textwidth]{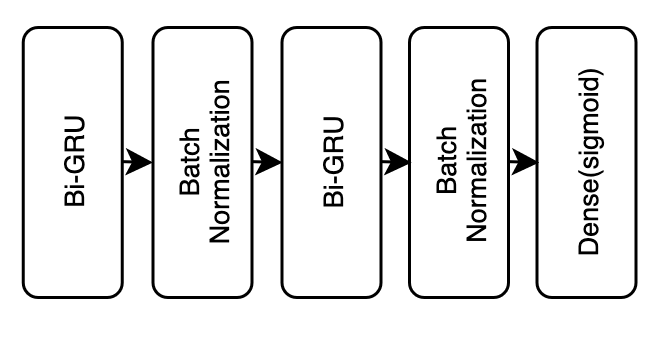}
    \caption{RNN-AE Encoders and Decoders architecture}
    \label{Fig:aeenc}
\end{figure}

\begin{figure}[!htb]
    \centering
    \subfigure[]{\includegraphics[width=0.4\textwidth]{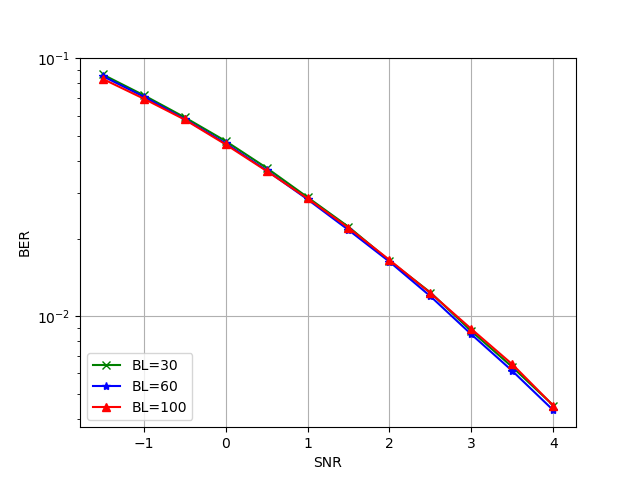}} 
    \subfigure[]{\includegraphics[width=0.4\textwidth]{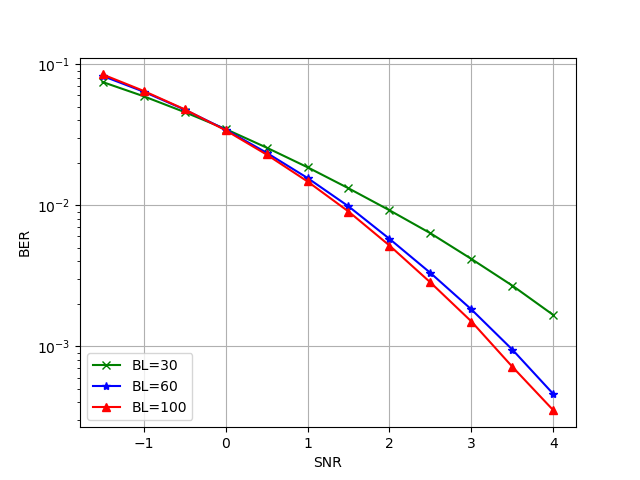}} 
    \caption{BER as a function of SNR evaluated at block lengths (BL): 30, 60, 100. (a) RNN-AE and (b) \tai. The reliability improves as the training/testing block length increases for DeepIC but not for RNN-AE.}
    \label{fig:blcg}
\end{figure}

It is well known in coding theory that as the block length of codewords increases, channel coding can achieve improved reliability \cite{berrou1993near}. To study if DeepIC exhibits such behavior, 
we compare the performances of \tai\  and RNN-AE when trained and tested with different training block lengths. In Figure~\ref{fig:blcg}, we plot the BER vs. SNR with train/test blocklengths varying from 30, 60, to 100 for (a) RNN-AE and (b) DeepIC at $h=0.8$. Surprisingly, we observe that 
as the block length increases, RNN-AE shows no improvement in BER, while \tai's BER decreases with a larger block length. We conclude that naively applying general purpose neural networks such as RNN to channel interference problems does not result in a better performance as we increase the block length. The encoders and decoders of RNN-AE share the same architecture, shown in Figure~\ref{Fig:aeenc}.

\subsection{\textbf {Degree of Interference}} 
\begin{figure}[!htb]
    \centering
    \subfigure[]{\includegraphics[width=0.4\textwidth]{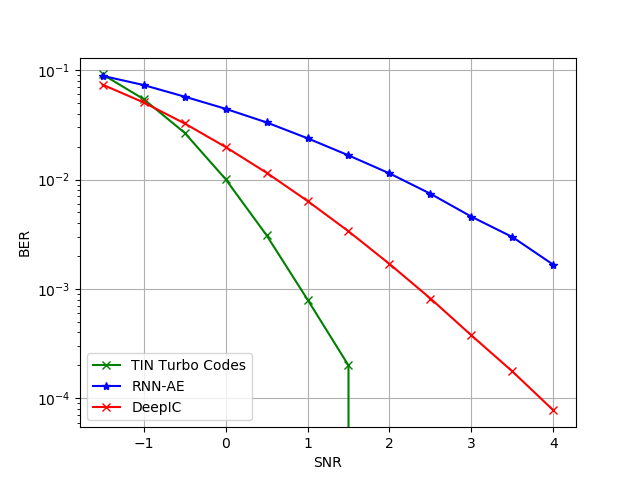}} \vspace{-.5em}
    \subfigure[]{\includegraphics[width=0.4\textwidth]{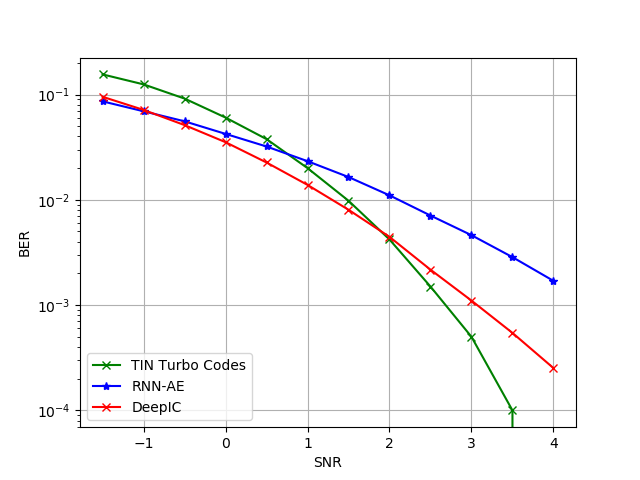}} 
    \caption{BER performance of Turbo Codes, RNN-AE and \tai\  for (a) h=0.3 (b) h=0.6. As the level of interference $h$ increases, DeepIC starts to outperform TIN.}
    \label{fig:0306}
\end{figure}

In Figure~\ref{Fig:fig1}, we considered $h=0.8$, which is a moderate interference regime. The strengths of the signal and the interference are similar, which makes coding for interference channels challenging. In this section, we vary $h$; we focus on smaller $h$'s to study if we can learn reliable codes for weak interference regimes. In particular, we consider $h=0.3$ and $h=0.6$, and the results are shown in Figure~\ref{fig:0306}.


While the superior performance of \tai\  is evident in the moderate interference regime ($h=0.8$), as the value of h decreases, \tai 's performance gets weaker when compared to classic Turbo codes with treating interference as noise (TIN). In such cases, training the DeepIC to ignore the interference and mimic Turbo codes would be optimal. We conjecture that DeepIC is not learning to mimic Turbo codes as DeepIC does not include an interleaver, which is the key component of Turbo codes. We will explore this intriguing problem in our future work. We discuss the role of interleaver in the following. 

\subsection{\textbf {Interleaved Structure of \tai\ }}
\begin{figure}[!htb]
    \centering
    \includegraphics[width=.4\textwidth]{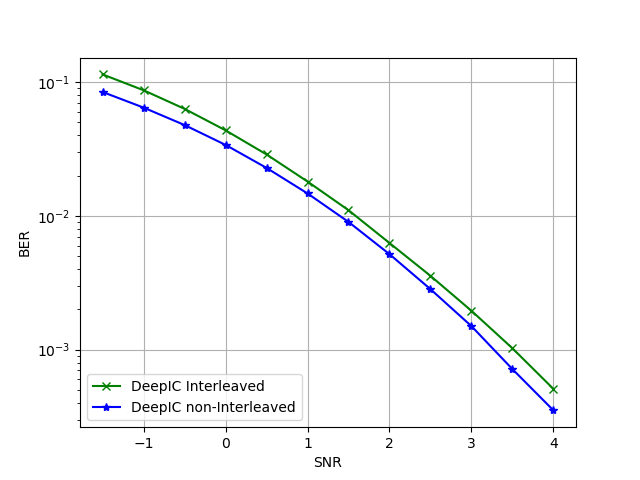}
\caption{BER performance of \tai\  with and without an interleaver: Interleaver marginally hurts the performance.}
    \label{Fig:fig7}
\end{figure}

As part of our experiments, we explore introducing an interleaver to both the encoding and decoding processes of \tai, which allows us to mimic Turbo autoencoders introduced in~\cite{Jiang2019}. At the encoder side, the input to CNN$_{3}$ of Figure~\ref{Fig:enc} would be shuffled by an interleaver $\pi$. Formally, $\bm{x_{3}}=\pi(\bm{b})$. At the decoder side, similar to before, the first decoder $f_{\phi_{1}}$ outputs the posterior $\bm{q_{1}}$. However,  $\bm{q_{1}}$ now goes through an interleaver $\pi$. The second decoder $f_{\phi_{2}}$ takes as input $\bm{y_{1}}$, $\bm{y_{3}}$ and $\pi(\bm{q_{1}})$, and outputs the second posterior $\bm{q_{2}}$, which passes through a de-interleaver $\pi^{-1}$. This process is repeated through multiple iterations. Figure~\ref{Fig:fig7} depicts the BER of \tai\  with and without an interleaver, trained on $h = 0.8$ and block length = 100. 

\noindent Although the gap is not significant, introducing an interleaver does hurt the training performance. This problem is an interesting research direction, and perhaps the interleaved structure might yield better results when applied with an even more complex training model. We will investigate this in our future work.

\subsection{\textbf{Interpretation}} 

\begin{figure}
    \centering
    \subfigure[]{\includegraphics[width=0.24\textwidth]{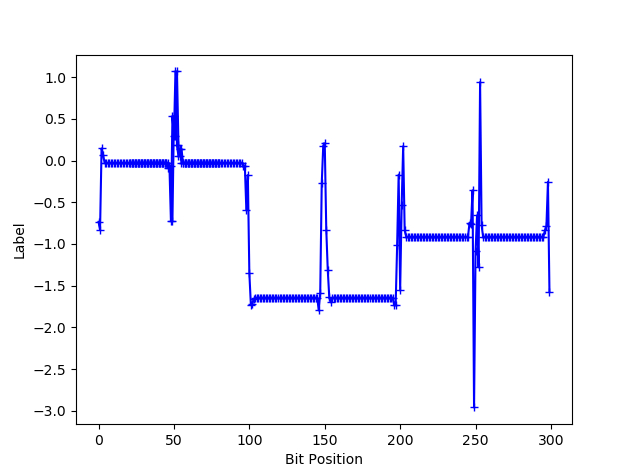}} 
    \subfigure[]{\includegraphics[width=0.24\textwidth]{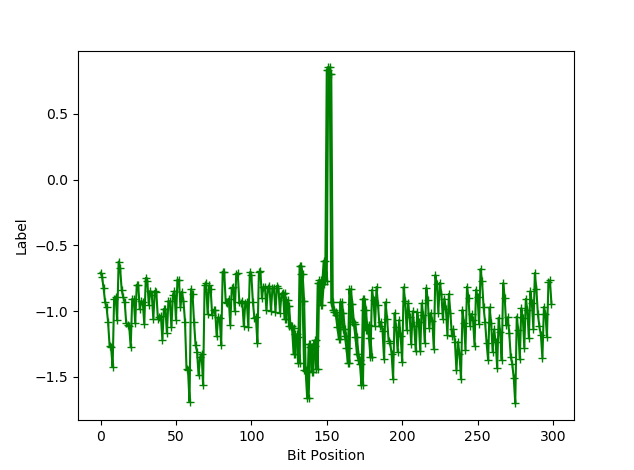}}
    \caption{Coded symbol vs. Bit position: Codeword representation of the message sequence $b^*$ of one of the two users for (a) DeepIC and (b) RNN-AE.} 
    \label{fig:fig10}
\end{figure}

A natural question is `what has DeepIC learned?' Interpreting the behavior of deep learning models is in general challenging, but we run several experiments to better understand the behavior of DeepIC. In the following, we present the perturbation analysis. 
We consider a special codeword $\bm{b^*}$, which consists of all '$0$' bits except for the middle bit which is a '$1$'. We feed $\bm{b^*}$ as an input to DeepIC and RNN-AE, and plot the encoded output of one of the two users, as shown in Figure~\ref{fig:fig10}.

By looking at the plots, we can verify the increased memory-dependence of the codeword when comparing DeepIC and RNN-AE, by noticing the duration of the perturbances caused by encoding the message bit '$1$'.  Note that the three peaks of Figure~\ref{fig:fig10} (a) correspond to the three concatinated encoders $g_{\theta_{1}}$, $g_{\theta_{2}}$ and $g_{\theta_{3}}$ presented in Figure~\ref{Fig:enc}, and the single peak of Figure~\ref{fig:fig10} (b) is due to the presence of just one encoder in RNN-AE's architecture.

\begin{figure}
    \centering
    \subfigure[]{\includegraphics[width=0.3\textwidth]{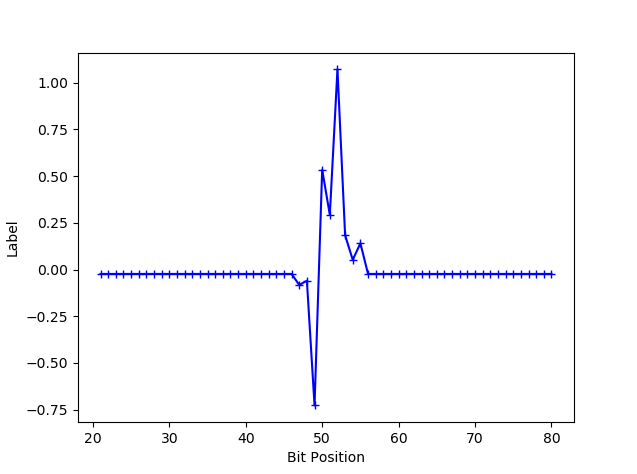}}
    \subfigure[]{\includegraphics[width=0.3\textwidth]{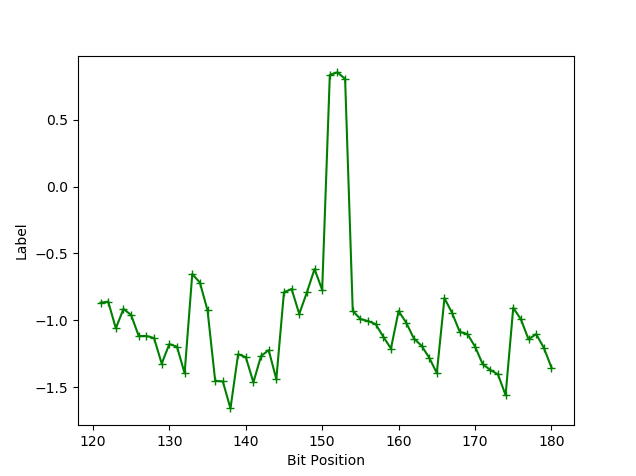}} 
    \caption{Coded symbol vs. Bit position: Codeword representation of the message sequence $b^*$ of one of the two users for (a) DeepIC zoomed on the first peak (b) RNN-AE zoomed on the middle peak.}
    \label{fig:fig10-zoom}
\end{figure}

In Figure~\ref{fig:fig10-zoom}, we show the zoomed-in plots for Figure~\ref{fig:fig10}. As is shown in the figure, the length of that perturbation is bigger for DeepIC than for RNN-AE. We see that 10 positions nearby the center are perturbed for DeepIC while only 3 positions nearby the center are perturbed for RNN-AE. This better explains the results described in Figure~\ref{fig:blcg}, and is one of the key reasons of the dominating performance of DeepIC when compared to other schemes, as we saw in Figure~\ref{Fig:fig1}.

\section{Conclusion and discussion}
\label{sec:conclusion}
In this paper, we show that properly designed and trained CNN codes with iterative decoders, which we call DeepIC, outperform the state-of-the-art by a significant margin on the challenging problem of communicating over two-user AWGN interference channels. 
%
While it had been shown in the literature~\cite{o2017introduction} that neural network based codes, in particular, feedforward networks, can outperform the time division for a very short block length, such as $K=4$ bits, this work is the first to show that such improvement is possible for longer blocklengths, such as $K=100$ bits.

\noindent\textbf{Learning codes for longer blocklengths.} As blocklength increases, outperforming the TD and TIN via deep learning becomes more challenging because TD and TIN implement the Turbo codes, the reliability of which improves as the blocklength increases. 
In order to outperform TD and TIN for longer blocklengths, neural codes need to learn to utilize memory in the encoding process. In other words, we want the $k$-th coded symbol to depend not only on $b_k$, but also on other $b_j$'s for $j \neq k$. We empirically show that DeepIC has a longer memory than other direct applications of existing neural architecture and training methodology in the literature. 
We also show that DeepIC outperforms all the baselines by a significant margin (upto 2dB gain vs. up to 1dB gain in~\cite{o2017introduction}). 

\noindent\textbf{Open problems.} There are several interesting open problems that extend from DeepIC. First is incorporating interleavers. While DeepIC has a longer memory than prior work, we believe its memory is still not as large as Turbo codes.  A natural way to resolve this limitation is to incorporate interleavers; harmoniously combining interleavers with DeepIC and devising appropriate training strategy is a very interesting open problem. Second, we mainly focused on a rate-1/3 setting. Learning codes for other rates are left as future work. 
Last, applying DeepIC on more complicated interference channels, including fading channels and channels with impediments, is another promising direction of future work.

\vspace{-.8em} 
 \medskip
 \small
 \bibliographystyle{IEEEtran}
 \bibliography{References}

\end{document}